\documentclass[12pt,preprint]{aastex}
\usepackage{CJK}
\shortauthors{Kwok \& Zhang} 
\shorttitle{PAH or MOAN}

\received{}
\usepackage{graphicx}
\begin{document}

\title{Unidentified Infrared Emission bands: PAHs or MAONs?}

\begin{CJK*}{Bg5}{bsmi}
\CJKtilde

\author{Sun Kwok (³¢·s), Yong Zhang (±iªa)}

\affil{Department of Physics, The University of Hong Kong, Pokfulam Road, Hong Kong, China}
\email{sunkwok@hku.hk}

\begin{abstract}

We suggest that the carrier of the unidentified infrared emission (UIE) bands is an amorphous carbonaceous solid with mixed aromatic/aliphatic structures, rather than  free-flying polycyclic aromatic hydrocarbon (PAH) molecules.  Through spectral fittings of the astronomical spectra of the UIE bands, we show that a significant amount of the energy is emitted by the aliphatic component, implying that aliphatic groups are an essential part of the chemical structure.  Arguments in favor of an amorphous, solid-state structure rather than a gas-phase molecule as a carrier of the UIE are also presented.

\end{abstract}


\section{Introduction}

Although the existence of organic grains in space has been speculated for some time \citep{hoy77,dul79}, the first serious identification of organic solid-state particles in the interstellar medium came after the discovery of the unidentified infrared emission (UIE) features in the planetary nebula NGC 7027 \citep{rus77}.  The UIE feature at 3.3 $\mu$m was first identified as the C--H stretching mode of aromatic compounds by  \citet{kna77}.  The possibility that the absorption bands seen at 3.4 and 6.9 $\mu$m  could arise from stretching and bending modes of complex hydrocarbons was raised by \citet{puetter79}.  The organic affiliation was extensively discussed by \citet{dul81} who assigned the 3.3 and 11.3 $\mu$m features to graphitic (aromatic) materials.  Subsequently, \citet{leg84} suggested that the UIE features arise from polycyclic aromatic hydrocarbon (PAH) molecules excited by single-photon heating.  In addition to the 3.3 and 11.3 $\mu$m features, they identified the 6.2 $\mu$m feature as due to aromatic C--C stretch, the 8.6 $\mu$m feature as the C--H in-plane bend, and the 11.3, 12.4 and 13.3 $\mu$m features as due to solo, duo and trio C--H out-of-plane bending modes.  The PAH hypothesis was extended by  \citet{all85} who calculated the vibrational spectrum of gas-phase chrysene and suggest possible matches to the UIE bands.  These works cumulated in two reviews in 1989 by \citet{all89} and  \citet{pug89} and the PAH hypothesis has become the dominant explanation for the UIE bands ever since. 

Although the literature has focused on the  3.3, 6.2, 7.7, 8.6, and 11.3 $\mu$m aromatic features, it should be recognized that the UIE phenomenon is much richer than these features.  Also present in astronomical spectra are emission features around 3.4 $\mu$m,   which arise from symmetric and asymmetric C--H stretching modes of methyl and methylene groups \citep{geballe1992}.  The bending modes of these groups also manifest themselves at 6.9 and 7.3 $\mu$m \citep{jou90,chiar2000, kwo01}. In addition, there are unidentified emission features at  15.8, 16.4, 17.4, 17.8, and 18.9 $\mu$m, which have been observed in proto-planetary nebulae \citep[PPN, ][]{kwo99}, reflection nebulae \citep{sel07}, and galaxies \citep{stu00}.

Most significantly, the UIE features are often accompanied by strong, broad emission plateaus features at 6--9, 10--15, and 15--20 $\mu$m.  The first two plateau features have been identified as superpositions of in-plane and out-of-plane bending modes emitted by a mixture of aliphatic side groups attached to aromatic rings \citep{kwo01}.  The 15--20 $\mu$m feature have been detected in young stellar objects, compact H{\small II} regions, and planetary nebulae, but are especially strong in some PPNs \citep{zhang2010}.  A possible explanation is that this broad feature arise from C--C--C in- and out-of-plane bending of aromatic rings \citep{van00}.



Although the UIE features are widely  attributed to the IR fluorescence of far-ultraviolet-pumped PAH molecules containing $\sim$50 C atoms \citep{tielens2008}, alternative carriers that have been proposed include hydrogenated amorphous carbon \citep[HAC,][]{buss1990, gad2012}, quenched carbonaceous composites  \citep[QCC,][]{sakata1987}, soot and carbon nanoparticles \citep{hu2008}, coal \citep{papoular1989}, and petroleum \citep{cataldo2002, cataldo2012}.  The arguments for a mixed aromatic/aliphatic organic nanoparticles (MAON) were recently summarized by \citet{kwok2011}.  In response to this suggestion, \citet[][herafter LD12]{li2012} made a modification to the PAH model to include a small number of methyl sidegroups to the PAH molecules (see Fig. 2. of LD12).   They also expand the definition of PAH beyond the usual chemical definition to include dehydrogenation, superhydrogenation, elemental substitutions, as well as minor aliphatic components. In spite of this revision of the PAH hypothesis, there remain fundamental differences between the PAH and MAON models.  Specifically (i) is the carrier of the UIE features a free-flying molecule or a solid?  (ii) are the chemical structural units regular with repeatable patterns, or amorphous with variable sizes and random orientations?  
(iii) is the carrier predominantly aromatic or a mixed $sp^2/sp^3$ structure?  In this paper, we try to address these questions.

\section{Can the PAH model fit the observed astronomical spectra?}
   
The UIE phenomenon consists of a family of well-defined discrete bands and broad plateau features, both sitting on top of a strong continuum.  Although the exact peak wavelengths, band profile, and relative strengths of the features vary from source to source, the general qualitative pattern of the phenomenon is remarkably consistent from circumstellar envelopes, to diffuse interstellar medium, to galaxies.  Since laboratory measurements have shown that no PAH emission spectrum has been able to reproduce the UIE spectrum with respect to either band positions or relative intensities \citep{cook1996, cook1998, wagner2000}, the PAH model has to appeal to a mixture of PAHs of different sizes, structures (compact, linear, branched) and ionization states, as well as artificial broad intrinsic line profiles to fit  the observed spectra.  The excellent fits to the observed spectra using this approach have been cited as evidence for the success of the PAH model \citep{tielens2008}. Here we will examine in detail how such fits are obtained.

The existing PAH model fits to the UIE spectrum can be divided into two categories.  A ``population synthesis'' \footnote{The population synthesis model referred to here are different from the population synthesis models of galaxies in that the stellar populations of galaxies are governed by initial mass function and other constraints whereas the relative abundance of PAH populations is entirely free.} model  \citep{all99, cami2011} uses the laboratory or theoretical spectral properties of a mixture of neutral and ionized PAHs and the abundance of each of constituents are then adjusted by a least-square-fitting method in order to produce a good fit.  Since the intrinsic profiles of the PAH molecules are narrow, artificial line widths have to be assigned to each of the features.  Specifically, \citet{cami2011} adopts FWHM of 13 cm$^{-1}$ as the profile width in the 6--10 $\mu$m range and 4 cm$^{-1}$ in the 10--14 $\mu$m range, corresponding to $\Delta\lambda$ of $\sim$0.08 $\mu$m at 8 $\mu$m and $\sim$0.06 $\mu$m at 12 $\mu$m.   The main problem with these models is the large number of free parameters and the non-uniqueness of the solutions.  

A variation of the above method is by single value decomposition method where the spectral-spatial data are fitted by a Monte Carlo search algorithm \citep{rap2005}.  The ISOCAM data of NGC 7023 are fitted by a mixture of cationic (PAH$^+$) and neutral PAH (PAH$^0$) plus a population of very small grains (PAH clusters).  This method is more specific in identifying the origin of the spectral features.  For example, the fitting can identify the 7.6 $\mu$m feature to be due to PAH, whereas the 7.8 $\mu$m feature is due to very small grains.

The second  approach is through empirical fitting.  These models  \citep{li2001, dra07} do not use actual PAH absorption profiles but use the observed astronomical UIE feature wavelengths as model PAH templates.  For the line profiles, \citet{li2001} assume linewidths with FWHM of 0.04, 0.20, 0.70, 0.40, and 0.20 $\mu$m for the UIE features at 3.3, 6.2, 7.7, 8.6, and 11.3 $\mu$m respectively.  Because the empirical fitting model starts from the observed astronomical spectra and not laboratory or theoretical PAH spectra, these assumed linewidths allow easy fittings to the observed spectra and the models only explore the effects of different grain size distributions and excitation conditions.  

\begin{figure}
\includegraphics[width=120mm]{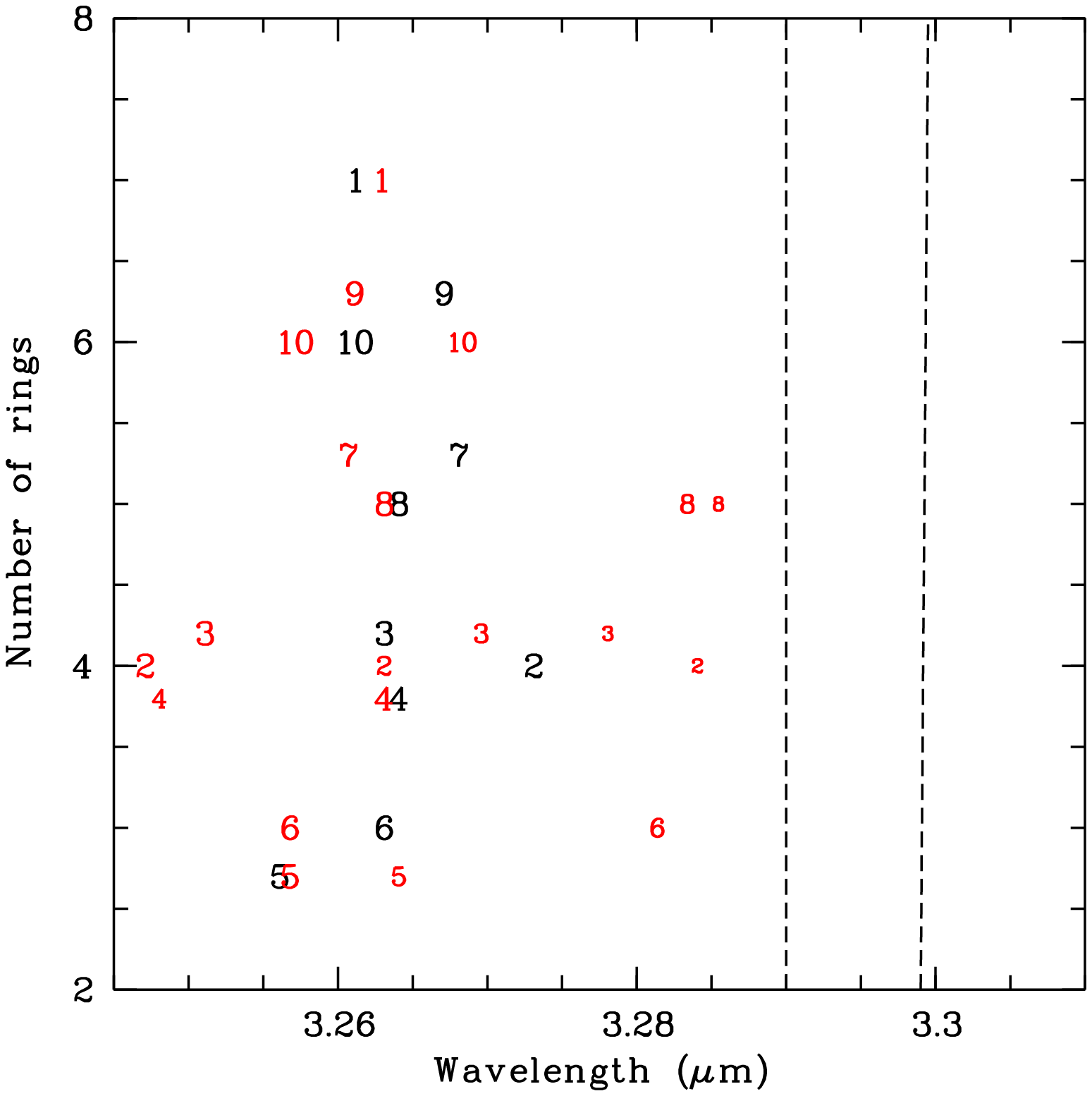}
\caption{The dashed lines represent the region within which the observed UIE band peak wavelength lies.  The black and red labels correspond to the experimental and theoretical values of the aromatic C--H mode of PAH molecules from the NASA Ames PAH database.  The molecules plotted are 1. coronene
2. naphthacene
3. chrysene
4. benzanthracene
5. phenanthrene
6. anthracene
7. benzo[e]pyrene
8. pentacene
9. benzoperylene
10. dibenzopyrene.  
When multiple peaks are observed, the size of the number is used to indicate the relative strength.  
\label{sakata}}
\end{figure}

The population synthesis models cover a wavelength range of 5-15 $\mu$m, leaving out the 3 $\mu$m region \citep[see discussion section of][]{peeters2011}. 
  \citet{sakata1990} noted that the actual C--H stretching modes of PAH molecules have center wavelengths shorter than 3.3 $\mu$m and  questioned  whether PAH molecules could fit the 3.3 $\mu$m UIE feature.  We have made use of the data in the NASA Ames PAH infrared spectroscopic database \citep{bauschlicher2010} from which the population synthesis models draw their data and plotted the central wavelengths of 18  PAH molecules. It is clear from Fig.~\ref{sakata} that the central wavelengths of the C--H stretching mode  lies shortward of the observed wavelength, consistent with the results previously shown by \citet{sakata1990}.  If the 3.3 $\mu$m feature arises from PAH molecules, then it cannot be due to neutral species. 

 \citet{joblin1995} consider the effects of anharmonic couplings and suggest that the astronomical 3.3 $\mu$m feature can be consistent with PAH molecules at high temperatures.  But the observed peak wavelengths of the 3.3 $\mu$m feature lie within a very narrow range in different circumstellar or interstellar environments.  This makes it unlikely that red shift by high temperatures be the general cause of this wavelength discrepancy.  
The aromatic C--H stretching mode of PAH has slightly different frequencies depending on how many of the edge H atoms are in ``bay'' or ``non-bay'' configurations \citep{baus09} and it is possible to separate these components by imaging spectroscopic observations \citep{candian2012}.  However, the peak wavelengths of the ``bay'' and ``non-bay''modes are at 3.17--3.27 and 3.26--3.27 $\mu$m, respectively, both  shortward of the astronomical wavelengths of the 3.28 and 3.30 $\mu$m components derived by \citet{candian2012} by fitting the asymmetric profile of the 3.3 $\mu$m feature.

The plateau features also represent a challenge to the PAH hypothesis.  
While the MAON model clearly identifies the 8 and 12 $\mu$m plateaus to be due to collective in-plane and out-of-plane bending modes of aliphatic side groups, LD12 can only assign these two plateau features vaguely to ``wings of C--C and C--H bands''.  If this is the case, then similar ``wings'' should be seen around the 3.3 $\mu$m feature but it is not generally observed.

\section{Gas-phase molecules or amorphous solid?}\label{solid}

\begin{figure}
\includegraphics[width=120mm]{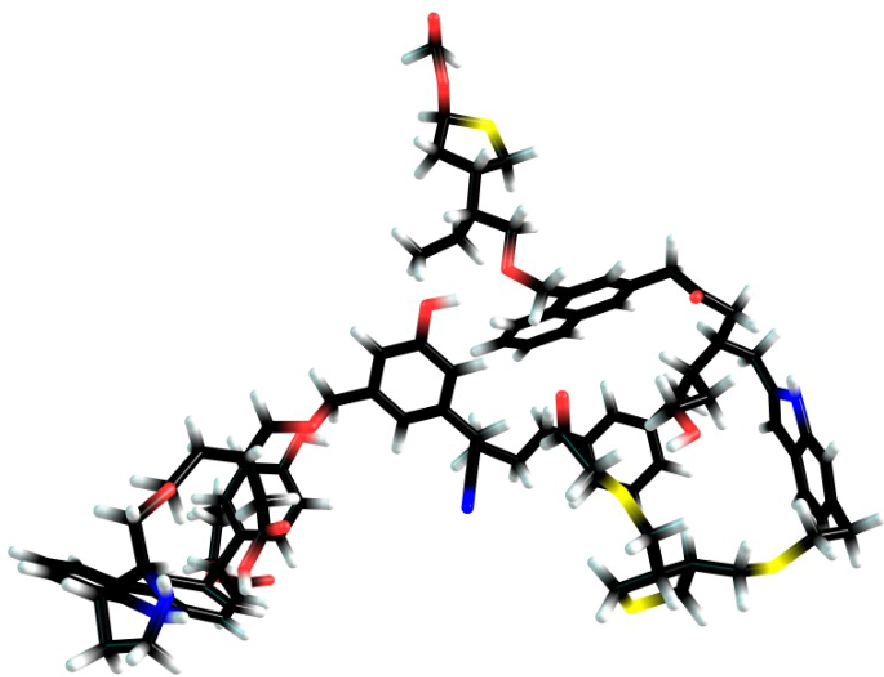}
\caption{A 3-D illustration of a possible partial structure of a MAON particle.  Carbon atoms are represented in black, hydrogen in light grey, sulphur in yellow, oxygen in red, and nitrogen in blue.  There are 101 C, 120 H, 14 O, 4 N, and 4 S atoms in this example.  The number of heavy elements have been intentionally exaggerated for the purpose of illustration.
\label{schematic}}
\end{figure}

In the MAON model, the carrier contains a mixture of aromatic and aliphatic units each with variable sizes and random orientations.  Such structure is very different from a PAH molecule where the structure, no matter how large, is regular with repeatable patterns.  The vibrational modes arising from an amorphous compound, whether they are due to aromatic or aliphatic units, will naturally produce broad emission profiles \citep{guillois1996, gad2012, cataldo2012}.  On the contrary, molecular bands are intrinsically narrow.  The spectral properties of a collection of gas-phase molecules are very different from those of an amorphous solid.  The PAH model assumes (and requires) that the molecular size to be small \citep[$<$50 C atoms,][]{tielens2008}, whereas the MAON model assumes the carrier to be a solid of hundreds or thousands of C atoms.  Another distinction between the PAH and MAON models is that PAH molecules have mostly plannar or curved layered structures whereas the MAON solids are fully 3 dimensional with disorganized   (Fig.~\ref{schematic}).  Therefore the PAH model, even in the modified form of LD12, is very different from the MAON model.  

PAH molecules are pure hydrocarbons.  However, in natural interstellar or circumstellar environments, other elements such as O, N, and S are abundantly present and can be expected to be incorporated into any carbonaceous compounds condensed from gas phase.  In Fig.~\ref{schematic}, we have illustrated how such impurities could be present in a carbonaceous particle. The purpose of this schematic is to illustrate the complexity of the chemical structure, as well as to show what kind of functional group (including those with heavy elements) may be contained in such structures.  However, the direct detection of such impurities would require higher sensitivity and spectral resolution than our existing  infrared spectroscopic facilities can offer.

We note that mixed aromatic/aliphatic structures are natural products of combustion. The first nucleation products of soot particles formed in flames have structures consisting of islands of aromatic rings linked by chains \citep{chung2011}.  Laboratory synthesized carbonaceous nanoparticles are found by chemical analysis to consist of networks of chains and rings \citep{hu2006}.  
Artificial synthesis of carbonaceous nanoparticles in the laboratory often results in disorganized materials with mixed aromatic/aliphatic structures.  The substances produced from vapor deposition is independent of energy injection techniques, which include laser ablation of graphite \citep{scott1996, mennella1999, jager2008}, laser pyrolysis of gases \citep{herlin1998}, arc discharge \citep{mennella2003},  microwave irradiation \citep{sakata1984, wada2009, god2011}, UV photolysis \citep{dartois2004}, and flame synthesis \citep{car2012}. 

Even with only H and C, amorphous solids have remarkable rich structures.  By introducing hydrogen into familiar forms of carbon such as graphite ($sp^2$) and diamond ($sp^3$), a variety of amorphous carbon-hydrogen alloys can be created. PAHs only represent the simplest form of hydrogen-carbon crystalline structures \citep{robertson2002}.   While it is likely that PAHs will exist in space, there is no reason to expect that it is the dominant hydrocarbon species.  In fact, the preference for PAH only represents a bias resulting from our previous lack of knowledge of the richness of hydrogen-carbon structures.

By varying the aromatic ($sp^2$) to aliphatic ($sp^3$) ratio and the H content, geometric structures of different long- and short-range order can be constructed.  These amorphous carbonaceous solids have optical properties that can easily reproduce the peak wavelengths and profiles of observed UIE features without the artificial assumptions and parameter fiddling as in the PAH model \citep{jones2012a}.

The fact that solids are present in the UIE band emission region is not disputed as most of the UIE bands are observed with strong underlying continuum, which can only be due to large (micron-size) solid-state particles.  

\section{Aromatic or mixed?}
\label{fitting}

One of the key question of the debate is the relative fractions of $sp^2/sp^3$ hybridization in the carrier of the UIE bands.  LD12 acknowledged that the UIE features probably do not arise from pure PAH molecules and  proposed a model of superhydrogenated methyl-PAHs as the carrier of the features.  We note that these superhydrogenated methyl-PAH is in fact similar to the synthetic carbonaceous material filmy QCC, a yellow-brown material made up of clusters of pure, methylated and ethylated PAHs \citep{sakata1987, wada2009}.  The major difference is that the LD12 model is gaseous and free-flying, whereas flimy QCC are solids.  Filmy QCC can be considered as an intermediate compound between PAHs and MAONs.

LD12 arrive at the conclusion that the aliphatic component is not important based the assumption that the 3.4 $\mu$m feature is the only manifestation of aliphatic structures.  
The 3.4 $\mu$m feature is due to stretching modes of methyl and methylene groups, two simplest examples of aliphatic chains that can attach to or link the aromatic rings.  After the first ring molecule (benzene) is formed from the acetylene, other rings can cluster together for form aromatic units.  However, it is likely that many other functional groups will attach to the rings.  These side groups can take many forms (--CH=CH$_2$, --CH=CH--, C=CH$_2$, C=C--H, etc) and some examples are shown in Figure 4 of \citet{kwo01}.  Such aliphatic groups  not necessarily attached to the rings and can also form their  own networks, as illustrated in Fig.~\ref{schematic}.

In a natural chemical synthesis, it is extremely unlikely that the aliphatic component would be limited to simple methyl and methylene groups.  In addition to the 3.4 $\mu$m feature,  this complex of side groups manifests themselves through in-plane and out-of-plane bending modes around 8 and 12 $\mu$m.  Therefore, to correctly assess the contributions of the aliphatic component, one has to include the plateau emissions. 

The 8 and 12 $\mu$m plateau features are found to be strong in sources where fullerene (C$_{60}$) is found \citep{ber2012, garcia2012}.  Since several such sources (Tc-1, SMP SMC 16, SMP LMC 56) have no observable aromatic features, this demonstrates that these plateau features are NOT aromatic in origin and therefore are definitely not related to PAHs, contrary to the explanation offered by LD12.   In fact, this observed correlation has led to the suggestion that the C$_{60}$ molecules are the break-down products of MAONs \citep{ber2012, garcia2012}.  Separately, \citet{mic2012} also suggest that the precursors of C$_{60}$ are clusters consisting of small aromatic islands linked by aliphatic bridges (called ``arophatics''), a structure simpler but similar to that of MAONs.  

In order to quantify the relative contributions from different components, we have performed spectral decomposition of the infrared spectra of several different kinds of sources exhibiting the UIE bands.
The spectral data for the objects are retrieved from the {\it Spitzer Space Telescope} and the {\it Infrared Space Observatory (ISO)} archives.  For the spectral decomposition, we used the IDL package PAHFIT originally developed to fit the {\it Spitzer} IRS spectra of nearby galaxies \citep{smi07}. The model spectra take into account the contributions from stellar continuum, thermal dust continuum, H$_2$ emission, atomic emission lines, the UIE features (both aromatic and aliphatic), and the plateau emission features.  For the cases of IRAS 21282+5050, 2MASS J06314796+0419381 and M82, atomic lines are present and they are not included in the fitting.

The optimal fitting to the observed spectra is achieved through the Levenberg-Marquardt least-squares algorithm.  A modified blackbody model {\it I$_\lambda$$\sim$$\lambda$$^{-a}$B$_\lambda$(T)}, where {\it B$_\lambda$(T)} is the blackbody function with a temperature {\it T} is adopted to simulate the dust thermal continuum. The aromatic, aliphatic, and plateau features are fitted with assumed Drude profiles 
\begin{equation}
I_\lambda =  \frac{I_0 \gamma^{2}}{(\lambda/\lambda_{0}  -  \lambda_{0}/\lambda)^{2} + \gamma^{2}}
\end{equation}
where $\lambda_0$ is the central wavelength, $I_0$ is the central intensity, and $\gamma$ is the fractional FWHM of each feature.

\begin{figure}
\includegraphics[width=120mm]{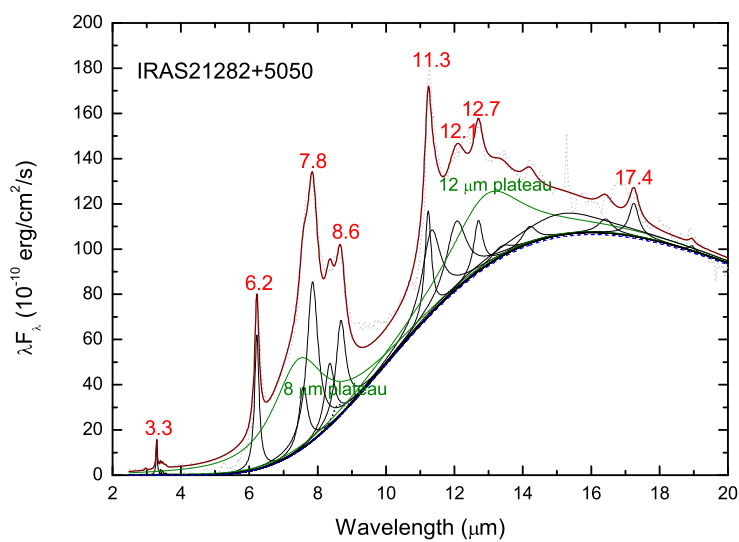}
\caption{A spectral decomposition of the UIE features of the young planetary nebula IRAS 21282+5050.  A series of discrete (black lines) and plateau features (green lines) above a continuum of the form $\lambda^{-2}B_\lambda$(150 K) (blue line) have been fitted to the observed data (light dotted line). The wavelengths of the UIE bands are labeled in red.   The spectra (5-20 $\mu$m) of IRAS 21282+5050 were taken from the {\it ISO} archive  and ground-based data from the Keck Observatory \citep[3-3.8 $\mu$m,][]{hrivnak2007}.  The Keck spectrum has been adjusted upward by a factor of 2.3 to line up with the {\it ISO} data.
\label{21282}}
\end{figure}

The infrared spectrum of IRAS 21282+5050 presented in Fig.~\ref{21282} are typical of that of carbon-rich PN \citep{stan2012}.  A strong infrared continuum originating from dust emission accounts for about 1/3 of the energy output of the nebulae with the remaining coming from the nebular gas component and the central star.  The dust continuum can be approximated by a blackbody of 150 K, modified by an emissivity index ($\alpha$) of $2$.   Of the $\sim1.5\times10^{-11}$ W m$^{-2}$ emitted between 3 and 20 $\mu$m, $\sim$65\% is emitted in this dust continuum  Although {\it ISO} LWS data exist for IRAS 21282+5050, the quality of the data is not good.  However, we can extend the Rayleigh-Jeans part of the modified blackbody to long wavelengths and obtain an estimate of the total flux between 2 and 200 $\mu$m of  2.1 $\times 10^{-11}$ W m$^{-2}$. 
Above this continuum are two broad plateau features around 8 and 12 $\mu$m.  From the spectral decomposition fittings shown in Fig.~\ref{21282}, the two plateau features together account for 17\% of the total flux between 2 and 20 $\mu$m.  On top of the plateau features are the UIE bands and their relative contributions are given in Table 1.

\begin{figure}
\includegraphics[width=120mm]{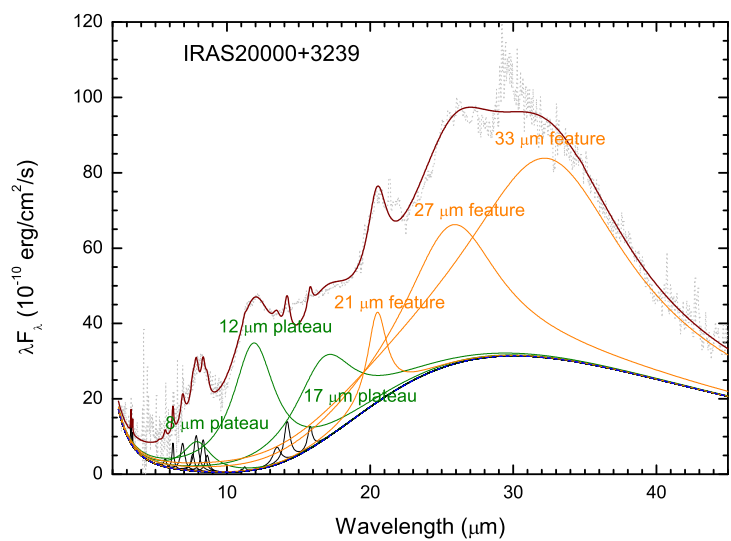}
\caption{A spectral decomposition of the UIE features of the proto-planetary nebula IRAS 20000+3239.  A series of discrete (black lines) and plateau features (green lines) above a continuum of the form $\lambda^{-2}B_\lambda$(80 K) (blue line) have been fitted to the observed data (light dotted line).   There are also two strong unidentified emission features at 21 and 30 $\mu$m and they are plotted in orange.  The spectra (2.4--45 $\mu$m) of IRAS 20000+3239 were taken from the {\it Spitzer} IRS  (4.8--19.5 $\mu$m) and {\it ISO} SWS (2.4--4.8 and 19.5--45 $\mu$m) observations.  
\label{20000}}
\end{figure}

A spectral decomposition of the infrared spectrum of the PPN IRAS 20000+3239 is shown in Fig.~\ref{20000}.  In addition to the UIE features, the strong unidentified emission features at 21 and 30 $\mu$m are also present \citep{hrivnak2009}. The broad 30 $\mu$m feature is fitted by two emission profiles at 25 and 33 $\mu$m.  It is clear that significant amount of energy is emitted in these two features and the strongest among the UIE features is the 12 $\mu$m plateau.

\begin{figure}
\includegraphics[width=120mm]{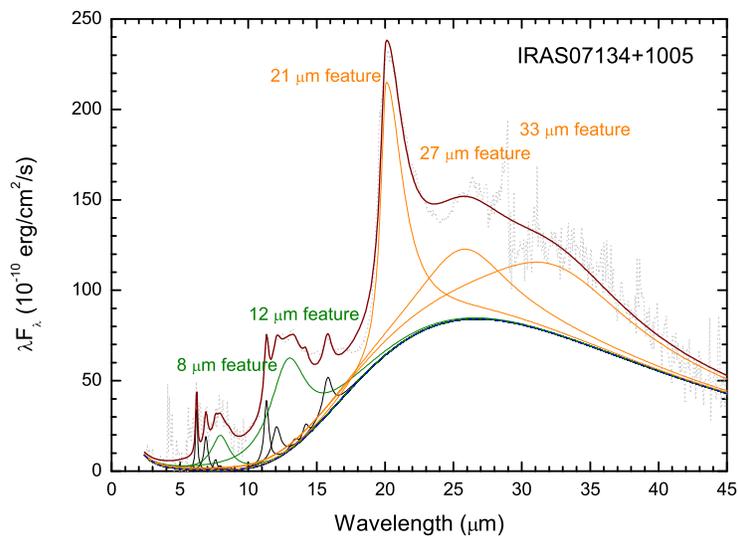}
\caption{A spectral decomposition of the UIE features of the proto-planetary nebula IRAS 07134+1005.  A series of discrete (black lines) and plateau features (green lines) above a continuum of the form $\lambda^{-2}B_\lambda$(90 K) (blue line) have been fitted to the observed data (light dotted line).  The unidentified 21 and 30 $\mu$m emission features are shown in orange.  
The spectra of IRAS 07134+1005 were taken from {\it Spitzer} IRS  (9.8--19.5 $\mu$m) and {\it ISO} SWS and LWS (2.4--9.8 $\mu$m and 19.5--200 $\mu$m) observations.
The rise towards the short wavelengths in the near infrared is due to photospheric continuum from the central star. 
\label{07134}}
\end{figure}

The 21 and 30 $\mu$m features are also seen in other carbon-rich PPNs, such as IRAS 07134+1005 (Fig.~\ref{07134}).  While the carrier of these two strong features are not yet identified, it is likely that they are carbonaceous as they are only observed in carbon-rich objects.  This suggests that carbon can be in many forms and do not reside exclusively in a single species such as PAHs.  

From the fitted dust continuum between 2 and 45 $\mu$m, we can extrapolate the dust continuum emission to 200 $\mu$m to estimate the total amount of energy emitted by IRAS 20000+3239 and IRAS 07134+1005.  These results are listed in Table 1.  The fraction of fluxes emitted by the aromatic and aliphatic components from these estimated total fluxes are also listed in Table 1.

\begin{figure}
\includegraphics[width=120mm]{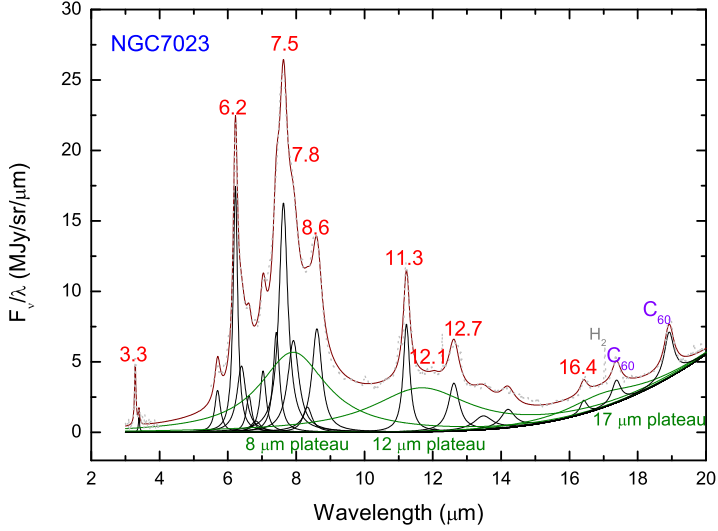}
\caption{A spectral decomposition of the UIE features of the reflection nebula NGC 7023.  Observational data are from the {\it Spitzer} IRS courtesy of K. Sellgren.  Data in the 3--4 $\mu$m region is from {\it ISOPHO}.  The underlying continuum is fitted by a modified blackbody ($\lambda^{-2}B_\lambda$(60 K)).  The marked features are UIE features and their respective contributions to the total spectrum are shown as spectral components under the total spectrum.  At the base of these features are three broad plateau features (green lines) centered at 8, 12 and 17 $\mu$m.  Because of the coolness of the central star ($T\sim 17\,000$ K), there are no atomic emission lines in this spectrum.  The only line present is the $v=0-0~ J=3-1$ 17.03 $\mu$m rotational line of molecular hydrogen.
The features at 17.4 and 18.9 $\mu$m have been suggested to originate from the C$_{60}$ molecule \citep{sellgren2010}.
\label{7023}}
\end{figure}

In the case of the reflection nebula NGC 7023, the continuum is considerably cooler and has only minor contribution to the total fluxes in the 2--20 $\mu$m region (Fig.~\ref{7023}).   The 7.7 $\mu$m feature is probably a blending of two peaks at 7.5 and 7.8 $\mu$m.  In addition to the 11.3 $\mu$m C--H out-of-plane bending mode, there are also features at 12.1 and 12.7 $\mu$m, which could be due to the duo and trio modes of the C--H out-of-plane bending.  On top of the 17 $\mu$m plateau, there are discrete features at 16.4 $\mu$m, 
the 18.9 $\mu$m features due to C$_{60}$, and the 17.4 $\mu$m band is a blend between aromatic and C$_{60}$ features \citep{sellgren2010}.

Although no far infrared spectral data exist for NGC 7023, {\it Herschel Space Observatory} 70, 160, 250 $\mu$m data suggest that the dust continuum component has a temperature of 25-30 K and the spectral energy distribution (SED) of the nebula peaks at a wavelength beyond 100 $\mu$m \citep{abergel2010, berne2012}.  This means that most of the continuum flux is emitted at wavelengths beyond 20 $\mu$m.  We have used the fitted dust continuum to extrapolate to 200 $\mu$m to estimate the total fluxes emitted.

\begin{figure}
\includegraphics[width=120mm]{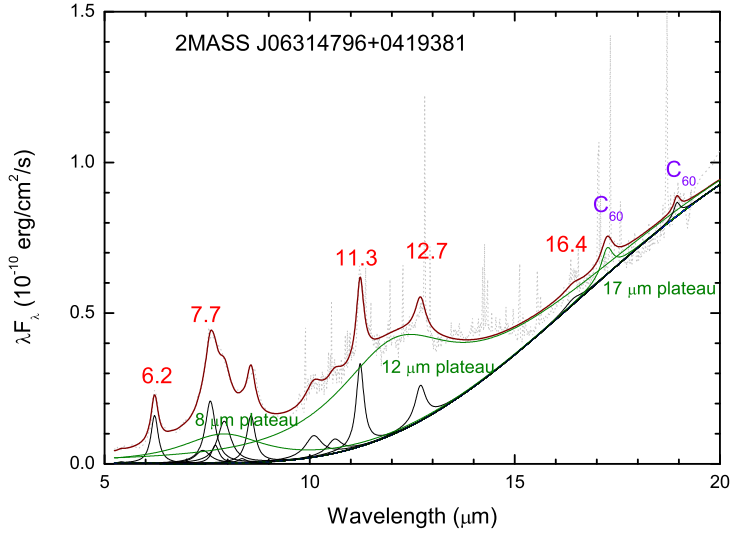}
\caption{A spectral decomposition of the UIE features of the YSO 2MASS J06314796+0419381.  Observational data are obtained from the {\it Spitzer} archive.   The underlying continuum is fitted by a modified blackbody ($\lambda^{-2}B_\lambda$(90 K)).  The marked features are UIE features and their respective contributions to the total spectrum are shown as spectral components under the total spectrum.  The narrow lines are due to atomic and molecular hydrogen lines and are not included in the fitting.  At the base of these features are three broad plateau features (green lines) centered at 8, 12 and 17 $\mu$m.  
The features at 17.4 and 18.9 $\mu$m have been suggested to originate from the C$_{60}$ molecule \citep{roberts2012}.  
\label{j063}}
\end{figure}

Figure \ref{j063} shows the infrared spectrum of the young stellar object (YSO) 2MASS J06314796+0419381.  Similar to PNs and PPNs, strong and broad plateau features coexist with UIE bands, both sitting on top of a underlying continuum.  Some atomic lines, molecular hydrogen lines, and C$_{60}$ bands can also be seen.   

\begin{figure}
\includegraphics[width=120mm]{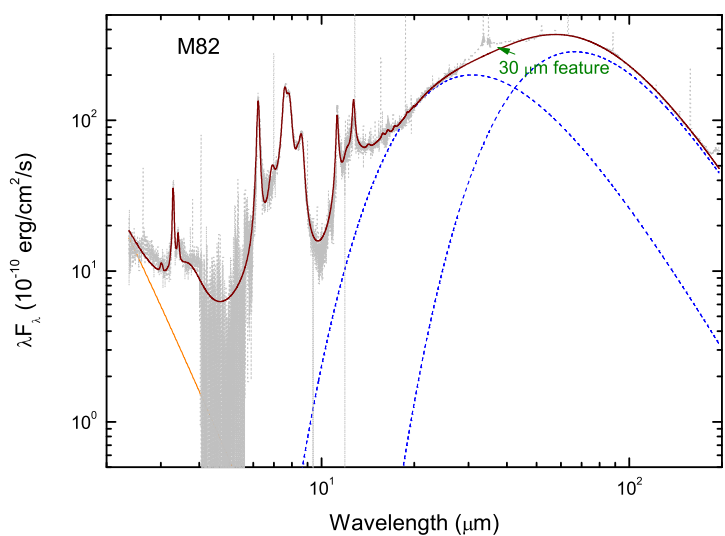}
\caption{The spectral energy distribution of the starburst galaxy M 82.  The spectra (shown as grey lines) of M 82 were taken from data obtained with the {\it Infrared Space Observatory} SWS and LWS instruments.  The UIE bands and  the unidentified 30 $\mu$m feature can be seen above the dust continuum. Two modified blackbodies in the form of  $\lambda^{-2}B_\lambda(T)$ (blue dashed line) are needed to fit the dust continuum.   Continuum emission from the central photoionized region is represented by a blackbody of 50000 K (shown as red dotted line).  Some of the strong atomic lines extend beyond  the vertical scale of the plot.  
\label{m82}}
\end{figure}

Figure~\ref{m82} shows the overall spectral distribution of the galaxy M 82.  Its SED is typical of other starburst and infrared luminous galaxies.  It is clear that most of the energy of the galaxy is emitted in the infrared through dust continuum emission.  However, the dust continuum is too broad to be fitted by a single temperature and we have fitted the continuum of M 82 by two modified blackbody components of temperatures 100 and 45 K.  Above these continua are several broad emission features.  The one around 30 $\mu$m is the unidentified 30-$\mu$m emission feature.  Below about 5 $\mu$m, the continuum rises again, due to contribution from the central accretion disk and photoionized region.  This continuum is approximated by a 50,000 K blackbody.  Since the near infrared represents the Rayleigh-Jeans tail of the blackbody, this continuum is not sensitive to the temperature used and can be fitted equally well, e.g., by a 100,000 K blackbody.  Since the SED represents the integrated light output of the galaxy, it clearly illustrates the importance of solid-state emission.  The strengths of the UIE bands also testify to  the prevalence of the organic matter in the galaxy. 

\begin{figure}
\includegraphics[width=120mm]{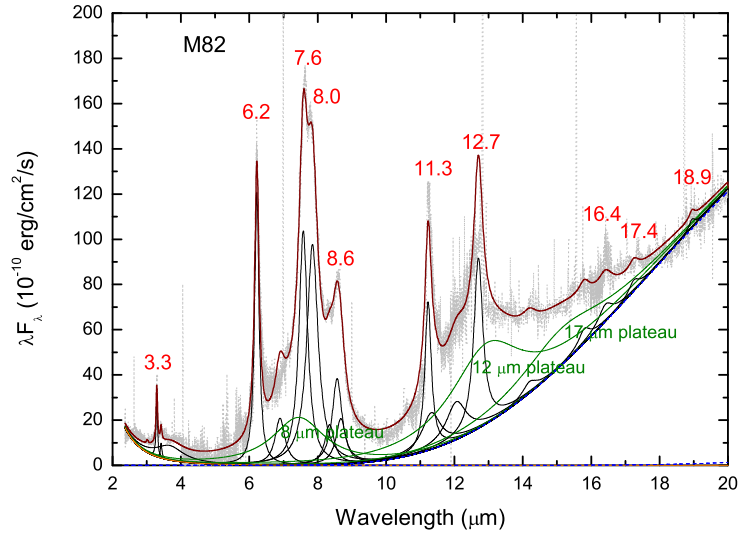}
\caption{Spectral decomposition of the UIE features in M 82.  A series of discrete (black line) and plateau (green line) features situated on top a continuum  (blue dashed line) have been fitted to the observed data (light dotted line).  The strong narrow features are atomic lines and are not included in the fitting.  
The wavelengths of some of the UIE features are marked in red.
\label{m82sws}}
\end{figure}

In order to better view the UIE features, a plot of the 3--20 $\mu$m region is shown in Fig.~\ref{m82sws}.  The prominence of the 8 $\mu$m (covering the wavelength ranges of 6--9 $\mu$m), 12 $\mu$m (10--15 $\mu$m), and 17 $\mu$m (15--18 $\mu$m) plateau features are clearly seen.  
A smaller plateau feature can be seen at 2-4 $\mu$m.
Above these plateau features, a host of UIE features are present \footnote{We note that the 17.4 and 18.9 $\mu$m features, which have been identified as due to C$_{60}$ in NGC 7023 and 2MASS J06314796+0419381, are also seen in the spectra of IRAS 21282+5050 (Fig.~\ref{21282}) and M 82 (Fig.~\ref{m82sws}).  It is possible that C$_{60}$ may also be present in these two objects.}.  The relative strengths of the features are given in Table 1.  

From the spectral fitting results in Table 1, we can see that the ratio of energy emitted by the aliphatic to that of energy emitted by the aromatic components are 1.0, 8.0, 4.6, 1.3, 3.4, and 0.7 for IRAS 21282+5050, IRAS 20000+3239, IRAS 07134+1005, NGC 7023, 2MASS J06314796+0419381 and M 82, respectively.  In computing these ratios, we have included the 3.3, 6.2, 7.7, 8.6, 11.3, 12.7, 13.4, 15.8, 16.4 and 18.9 $\mu$m features as aromatic, and 3.4, 6.9 $\mu$m features and the 8 and 12 $\mu$m plateau features as aliphatic.  The aliphatic component is clearly present and cannot be ignored in any chemical structure of the UIE carriers.  However, to translate the above intensity ratios to abundance ratios will require knowledge of the relative band strengths.  In general, the aliphatic transitions are stronger than aromatic transitions \citep{dartois2007}, so the relative aliphatic to aromatic abundance ratios (e.g., the 3.4 to 3.3 $\mu$m features ratio) are likely to be smaller than the values given above.  Since the observed plateau features are the result of unresolved blends of many modes from different aliphatic units, a derivation of their abundance is extremely difficult.

Evidence for the aliphatic structures is not limited to the 3.4, 6.9 $\mu$m and the plateau features.  The 6.2 $\mu$m feature is known to vary in peak wavelengths from PPN to PN.   By comparison with soot samples with known chemical structures, this variation is identified as a tracer of aromatic to aliphatic ratio \citep{pino2008}.  When the carrier is mainly aromatic, the band position is at 6.2 $\mu$m, but it shifts to 6.3 $\mu$m when the compound is mainly aliphatic.  In fact, the spectral evolution from PPN to PN provides powerful constraints on the evolution of the chemical structure of the carrier.  It is well known that thermal heating destroys aliphatic bonds \citep{sakata1990} and the change in the strengthening of the aromatic to aliphatic ratios from PPN to PN is likely the result of photochemistry \citep{kwo01}.  The strengths of the aromatic features is the result of bond migrations and cyclization of an originally complex carbonaceous compound.  This scenario is very different, and matches much better  observations, than the modified PAH hypothesis with pure PAH molecules picking up a few methyl groups along the way.

\section{Excitation mechanism}

In addition to the aromatic and free-flying molecular structures, another key element of the PAH hypothesis is excitation of the UIE features by single-photon transient heating mechanism.  As we see in Section \ref{fitting},  most of the energy in PN, PPN, reflection nebulae and starburst galaxies  is emitted in the continuum component.  The strong continuum radiation extends to far infrared and submmilimeter wavelengths, and the emission efficiency dependence on $\lambda/a$ ($a$ being the grain size) requires the existence of micron-size dust particles.  
Even in the diffuse ISM, a strong correlation exists between the strengths of the UIE features and the underlying continuum, suggesting that the UIE carriers and the large dust grains in the ISM are tightly related \citep{kah2003, sakon2004}.  \citet{kah2003} even suggest that these two components share a common heating mechanism as well as formation and evolution histories.  
The continuum is therefore an integral part of the UIE phenomenon and needs to be considered together.  

In PN and PPN, we know that both UIE and dust continuum components are formed in the circumstellar outflows and the central stars represent the only ultimate energy source for heating.  The dust continuum component is certainly heated by visible light from the central stars.  For starburst galaxies, the warm dust component is heated by the central photoionized region.  In PN and starburst galaxies, diffuse Lyman $\alpha$ may also contribute to the heating of the dust continuum \citep{kwok2007}.  In the case of PPNs, where there is no UV radiation background, the UIE features cannot rely on single-, high-energy, photon for excitation and must draw the energy from either the stellar radiation field or some other \citep[e.g., chemical,][]{dul2011, pap2012} energy sources.  In order to avoid the problem of neutral PAH molecules having weak absorption in the visible region, ionized PAHs are introduced as possible sources of absorption \citep[LD12,][]{mat2005}.   
Since PAH molecules have low ionization potential, given an adequate background UV radiation, they can be ionized in diffuse interstellar medium or photo-dissociation regions.  However, it is unlikely that PAHs will be extensively ionized in PPNs or reflection nebulae where the density is high and UV background is negligible.    




\section{Why is no PAH molecule detected?}

Since the UIE features are widely observed throughout the Galaxy and represent up to 20\% of total energy output of starburst galaxies \citep{smi07}, the carrier of these features must be extremely common and abundant.  PAH have very strong absorption in the UV (4-10 eV, or 0.12-0.3 $\mu$m).  If the abundance of PAH is high enough  to emit strongly in the infrared, it is difficult to understand why the same molecules in the ground electronic state will not cause absorption features in the UV \citep{mathis1998}.  Sensitive searches have been made in the UV with the {\it Hubble Space Telescope} and the {\it Very Large Telescope} but no detection was made, putting very low ($10^{-10}-10^{-8}$) upper limits on the abundance of PAH molecules in the diffuse ISM \citep{clayton2003, salama2011, gredel2011}.   The absence of UV absorption features of the diffuse ISM is in fact a strong argument against a molecular carrier.

The explanation offered why no specific PAH molecule has yet been detected in space is that there are a large varieties of PAH molecules, and no individual species is abundant enough to be seen.  If a collection of PAH mixture can produce strong, discrete infrared features, why a collection of PAH mixture does not produce strong, discrete UV absorption features?  LD12 suggest that  a collection of large PAH molecules may be responsible for the 220 nm feature.  However, PAH molecules have different UV absorption peaks but the central wavelengths of the 220 nm feature is remarkably constant, which is more consistent with absorption expected from an amorphous solid \citep{men1996, gad2011}.  For this wavelength invariance of the 220 nm feature to be explained by a population of diverse PAH molecules would require remarkable coincidence along many different lines of sight in the Galaxy.  

PAH molecules have well-defined rotational and vibrational signatures and  should be detectable in the microwave and infrared spectral regions if they are the carrier of the UIE  bands. Some symmetric PAH molecules have a few strong vibrational lines and should be detectable.  For example, coronene has strong lines at 18, 26, and 66 $\mu$m \citep{mat2009}.  The highly symmetric PAH, Corannulene (C$_{20}$H$_{10}$), has a bowl-like shape and has a large  dipole moment (2.07 debye) along its symmetric axis.  Its rotational spectrum has been measured in the laboratory \citep{lovas2005}.  The absence of these species poses severe limitations on the PAH hypothesis.


\section{Conclusions}

The UIE phenomenon is a complex one consisting of a variety of vibrational modes of aromatic and aliphatic compounds.  The carrier of the UIE bands is also likely to be chemically related to other co-existing spectral phenomena such as C$_{60}$ and 21 and 30 $\mu$m emission features as these spectral features are  very likely to be manifestation of carbonaceous  compounds.  The PAH hypothesis has served an useful purpose to stimulate the astronomical community's interest in the chemical structure of organics in the Universe for the past 25 years.  However, our understanding of carbonaceous organic solids has progressed a great deal since then. We now realize  that much more complex organic compounds can be produced by carbon and hydrogen.  Extrapolations from laboratory simulations suggest that amorphous carbonaceous solids can also form under circumstellar conditions.  Moreover, they have optical properties that can naturally produce the observed UIE bands whereas the PAH hypothesis has to appeal to hydrogenation, ionization, and other artificial modifications in order to reconcile with observational constraints.  Recognizing that the pure C--H carbonaceous compounds such as HAC and QCC represent an improvement over the PAH model, the MAON model takes a step further to incorporate impurities such as O, S, and N, as well as including  extensive aliphatic networks.  

We offer the following scenario for the origin of complex organic compounds in space.  After the nucleosynthesis of the element carbon and its dredge up to the surface in asymptotic giant branch stars, a variety of organic molecules  are synthesized in the stellar winds from these stars.  Among the over 70 molecular species detected in circumstellar envelopes include simple hydrocarbons in the forms of chains and rings.  Laboratory simulations of circumstellar conditions suggest that carbonaceous compounds of mixed aromatic/aliphatic structures can be created in the form of amorphous nanoparticles.  These complex organics manifest themselves through vibrational stretching and bending mode radiations, first directly observable in the PPN phase.  After the onset of UV radiation from the now hotter central star in the PN phase, the structure undergoes transformation as the result of photochemistry.  These MAON particles could also serve as sources of C$_{60}$ and carriers of the 21 and 30 $\mu$m features.  These complex organics are ejected into the interstellar medium and are spread throughout the Galaxy.  They can be observed in the diffuse ISM through both emission and absorption spectroscopy.  Continued UV processing could lead to further aromatization of the material.  Their wide presence is confirmed by observations in the integrated spectra of external galaxies.  Whether these MAON particles are related to the interstellar 220 nm feature and the diffuse interstellar bands is an open question in need of further studies.

The study of the UIE phenomenon has greatly broadened the  field of condensed matter astrophysics.  In addition to oxygen-based silicate grains, we have a large population of carbonaceous grains in the Universe.  Through a combined study between laboratory simulations and astronomical observations, we have come to appreciate the complexity of organic matter in space.

{\flushleft \bf Acknowledgment~}
We thank SeyedAbdolreza Sadjadi for producing Figure 2 and Alan Tokunaga for helpful discussions.  We would also like to thank an anonymous referee for his/her helpful comments.  
This work was partially supported by the Research Grants Council of the Hong Kong Special Administrative Region, China (project no. HKU 7031/10P.).

\clearpage

%

\clearpage

\begin{deluxetable}{lrrrrcrrrrr}
\tablecolumns{11} 
\tablewidth{0pc} 
\tablecaption{Strengths of the UIE discrete and plateau features} 
\startdata
\cutinhead{Aromatic features (\%)} 
$\lambda$ ($\mu$m) & 3.3 & 6.2 & 7.7\tablenotemark{a} & 8.6 & 11.3 & 12.7 & 13.4 & 15.8 & 16.4 & 18.9\tablenotemark{b}\\
\tableline
I21282 (2--20) & 0.21 & 1.46 & 8.47 & 1.89 & 2.89 & 0.77 & 0.31 & 0.20 & 8.23 &0.04 \\
I20000 (2--20) &0.25&0.49& 2.9& 0.51  &0.071& \nodata & \nodata & 0.43 &\nodata & \nodata \\
I07134 (2--20) & \nodata & 2.15 & 0.82 & 0.15 & 2.69& \nodata & 0.43 & 2.25 & \nodata &\nodata\\
N 7023 (2--20) & 1.35 & 8.4 & 29.3 & 7.9 & 6.1 & 2.30 & 0.02 & 0.11 & 0.87 & 0.86\\
J063 (2-20)&\nodata &1.80 &6.4 &1.51 & 1.78 & 0.95 & \nodata & \nodata & 0.18 & 0.14\\
M 82 (2--20)& 0.66 & 5.26 & 15.8 & 4.05 & 3.94 & 3.29& \nodata & 0.28 & 0.39 & 0.06\\
\tableline
I21282 (2--200) & 0.12 &0.84 &4.89 &1.09 &1.67 &0.44 &0.18 &0.12 &4.47 &0.02\\
I20000 (2-200) & 0.086 & 0.17 & 0.98 & 0.18 & 0.024  & \nodata & \nodata & 0.15 & \nodata &\nodata\\
I07134 (2--200) & \nodata & 0.80 & 0.30 & 0.06 & 1.00 & \nodata &0.16 &0.84 &\nodata \\
N 7023 (2--200)& 0.10 & 0.64 & 2.24 & 0.60 & 0.46 & 0.18 &0.002 &0.01 &0.07 &0.07\\
J063 (2--200)&\nodata &0.32 & 1.1 & 0.27 & 0.32 & 0.17 & \nodata & \nodata &0.03 & 0.03\\
M 82 (2--200)& 0.09 & 0.73 & 2.18 & 0.56 & 0.54 &0.45 &  \nodata& 0.038 & 0.054 & 0.008\\

\cutinhead{Aliphatic, plateau, and unidentified features (\%)}
$\lambda$ ($\mu$m) & 3.4 & 6.9 &  & & 8 & 12 & 17 &  & 21 &30 \\
\tableline
I21282 (2--20)        & 0.09& \nodata & \nodata & \nodata & 7.69 & 8.92 & \nodata &  &\nodata  & \nodata\\
I20000 (2--20) &   0.13  &  1.0    &       &     &  6.8    &  29.2    &  19.0       &         &    1.8     &11.0 \\
I07134 (2--20)   &\nodata & 2.05 & && 8.6 & 28.2 & \nodata && 8.3  &12.3\\
N 7023 (2--20) & 0.22 & 2.0\tablenotemark{c}  & \nodata & \nodata & 29.3 & 16.1 & 2.7 & \nodata & \nodata & \nodata\\
J063 (2--20)& \nodata & \nodata  &  &  & 10.0 & 32.7 &\nodata & \nodata & \nodata & \nodata\\
M 82 (2-20)& 0.18 & 1.81& \nodata & \nodata & 9.1  &12.5 & 5.7 & \nodata & \nodata & \nodata\\
\tableline
I21282 (2--200) &0.05  &\nodata &&& 4.44 & 5.15 &\nodata  & & \nodata &\nodata\\
I20000 (2--200) &0.045 &0.35&\nodata & \nodata & 2.32 &10.0 & 6.54 &  & 1.9 & 31.8\\
I07134 (2--200)  &\nodata & 0.76 & && 3.2 & 10.5 &\nodata && 6.2 &21.9\\
NGC 7023 (2--200)&0.03 &  0.23 &&& 3.4 &1.9 & 0.3& & \nodata &\nodata\\
J063 (2--200)& \nodata & \nodata  & \nodata & \nodata & 1.80 & 5.89 &\nodata & \nodata & \nodata & \nodata\\
M 82 (2-200) & 0.024 & 0.25 & \nodata & \nodata & 1.3 & 1.72 & 0.78 & \nodata & \nodata & \nodata\\

\cutinhead{Continuum}
\colhead{}    &  \multicolumn{3}{c}{\% of total flux} &  \multicolumn{4}{c}{Temperature (K)} & 
\multicolumn{3}{c}{$\alpha$}\\
\tableline
I21282 (2--20)    &      & 64.9 &  & \nodata & \nodata & 150  & \nodata &         & 2 & \\
I21282 (2--200)  &      & 80.5 &  & \nodata & \nodata & 150  & \nodata &         & 2 & \\
I20000 (2--20)  &      & 7.2  &   &        &         &80    &         &          & 2 & \\
I20000 (2--200)&       & 26.6 &   &         &         &80   &         &         & 2 &\\
I07134 (2--20)  &      & 23.7  &   &        &         &90    &         &          & 2 & \\
I07134 (2--200)&       & 46.0 &   &         &         &90   &         &         & 2 &\\
NGC 7023 (2--20)& \nodata & 5.9 & \nodata & \nodata & \nodata & 60  &\nodata & \nodata & 2 & \nodata\\
NGC 7023 (2--200)& \nodata & 89.4 & \nodata & \nodata & \nodata & 60  &\nodata & \nodata & 2 & \nodata\\
J063 (2--20)& \nodata & 49.9 & \nodata & \nodata & \nodata & 90  &\nodata & \nodata & 2 & \nodata\\
J063 (2--200)& \nodata & 90.7 & \nodata & \nodata & \nodata & 90  &\nodata & \nodata & 2 & \nodata\\
M 82 (warm) & \nodata & 37.4 & \nodata & \nodata & \nodata & 100  & \nodata & \nodata & 0.7 & \nodata\\
M 82 (cold) & \nodata & 50.3 & \nodata & \nodata & \nodata & 45  & \nodata & \nodata & 0.85 & \nodata\\

\cutinhead{Total flux (W m$^{-2}$)\tablenotemark{d}\tablenotemark{e}}
I21282 (2--20)  &    &         & \nodata & \nodata & 1.5(-11)  & \nodata & \nodata & \nodata & &           \\
I21282 (2--200)   &    &         & \nodata & \nodata & 2.1(-11)  & \nodata & \nodata & \nodata & &           \\
I20000 (2--20)&    &         &         &         & 5.4(-12) &           &         &         &         &  \\
I20000 (2--200)&    &         &         &         & 1.6(-11) &           &         &         &         &  \\
I07134 (2--20)&    &         &         &         & 7.1(-12) &           &         &         &         &  \\
I07134 (2--200)&    &         &         &         & 1.9(-11) &           &         &         &         &  \\
J063 (2--20)& \nodata & \nodata & \nodata & \nodata & 4.9(-14)  & \nodata & \nodata & \nodata & &\\
J063 (2--200)& \nodata & \nodata & \nodata & \nodata & 2.7(-13)  & \nodata & \nodata & \nodata & &\\
M 82 (2--20)& \nodata & \nodata & \nodata & \nodata  & 9.0(-12)   & \nodata & \nodata & \nodata & &\\
M 82 (2--200)& \nodata & \nodata & \nodata &\nodata  & 6.5(-11)  & \nodata & \nodata & \nodata & &\\

\enddata

\tablenotetext{a}{This column represents the total strengths of features at 7.6, 7.8, and 8.3 $\mu$m.}
\tablenotetext{b}{The 18.9 $\mu$m feature can be due to C$_{60}$}.
\tablenotetext{c}{The central wavelength of this feature is at 6.7 $\mu$m.}
\tablenotetext{d}{With the exception of M82 where ISO LWS data are available, the total fluxes between 2--200 $\mu$m are estimated from extrapolation of the modified blackbody dust continuum.}
\tablenotetext{e}{Total flux is not available for NGC 7023.}
 
\end{deluxetable}

\end{CJK*}

\end{document}